\documentclass[11pt]{article}
\usepackage{feynmf}
\renewcommand{\thefootnote}{\fnsymbol{footnote}}
\newcounter{line}  
\def\ie{\hbox{\it i.e.}{}}

\def\nn{\hspace{2mm}}
\def\sss{\scriptscriptstyle}
\def\MeV{\mbox{\rm MeV}}
\def\GeV{\mbox{\rm GeV}}
\def\TeV{\mbox{\rm TeV}}

\def\sleq{\raisebox{-.6ex}{${\textstyle\stackrel{<}{\sim}}$}}
\def\sgeq{\raisebox{-.6ex}{${\textstyle\stackrel{>}{\sim}}$}}

\def\Hat#1{\widehat{#1}}

\def\sVEV#1{\left\langle #1\right\rangle}

\def\abs#1{\left| #1\right|} 

\def\CP{C\!\!\:P}
\def\AGUT{{}\;\;\raisebox{.9ex}{$\times$}\raisebox{-.5ex}%
{$\!\!\!\!\!\!\!\!\sss i=1,2,3$} \,(SMG_i \times U(1)_{\sss B-L,i})}

\begin{document}
\begin{titlepage}
\hfill
\vbox{
    \halign{#\hfil        \cr
           NBI-HE-00-48    \cr
           hep-ph/0101307 \cr
           }
      }
\vspace*{20mm}
\begin{center}
{\Large{\bf  Baryogenesis via lepton number violation in Anti-GUT model}\\}
\vspace*{15mm}
\vspace*{1mm}
{\ H. B. Nielsen}\footnote[3]{E-mail: hbech@alf.nbi.dk}
and {\ Y. Takanishi}\footnote[4]{E-mail: yasutaka@nbi.dk}

\vspace*{1cm} 
%\vskip .5cm
{\it The Niels Bohr Institute,\\
Blegdamsvej 17, DK-2100 Copenhagen {\O}, Denmark}\\

\vspace*{.5cm}
%\maketitle
\end{center}

\begin{abstract}
We study the baryogenesis via lepton number violation
in the model of Anti-GUT. The origin of the baryogenesis 
is the existence of right-handed Majorana neutrinos 
which decay in a $C$, $CP$ and lepton number violation way.
The baryon number asymmetry is calculated in the extended 
Anti-GUT model which is only able to predict order of 
magnitude-wise. We predicted baryon number to entropy ratio,
$Y_B=1.46{+5.87\atop-1.17}\times10^{-11}$, and this result 
agrees with experimental values very well.

\vskip 2mm \noindent\
\vskip -3mm \noindent\
\end{abstract}
\vskip 20mm

%\vskip 4cm
January 2001
\end{titlepage}

\newpage
\renewcommand{\thefootnote}{\arabic{footnote}}
\setcounter{footnote}{0}
\setcounter{page}{2}
%%%%%%%%%%%%%%%%%%%%%%%%%%%%%%%%%%%%%%%%%%%%%%%%%%%%%%%
\section{Introduction}

The evidence of the neutrino masses and 
its mixing angles from the atmospheric and solar
neutrino experiments 
\cite{SK1,SK2,Talk1,Talk2,Chlorine,SAGE,GALLEX,BKS,Valle} indicates a method 
to solve a challenging question in cosmology and particle physics:
namely, baryon asymmetry of the Universe. This asymmetry cannot be 
explained with the ``pure'' Standard Model (SM) -- with negligible $B-L$ 
asymmetry -- to the phenomenologically right magnitude of 
baryogenesis.  
In fact, the electroweak phase transition scenario for baryogenesis 
does not work very well in the frame of the SM with 
presently known lower bound of the SM Higgs 
mass \cite{wpt}. Even for the minimal supersymmetric 
standard model (MSSM) this scenario is strongly disfavoured by 
baryon number wash-out at the electroweak phase 
transition \cite{colin}. More detailed analyses of the 
MSSM \cite{LCMQSW} have been made and it is claimed that the
MSSM is just consistent with baryogenesis in a very
restricted region of parameter space requiring the
right-handed stop to be lighter than the top quark
and the left-handed stop heavier than 1~$\TeV$.    

The model which we investigate in the present article -- the extended 
Anti-GUT model -- has as one of its characteristics that it coincides 
with the pure SM (at least) for energy scales below the see-saw neutrino 
scale, and so we have in this model no way to get the phenomenological 
baryon number unless we already have an $B-L$ asymmetry prior to the 
weak epoch.

However in the SM the neutrinos are massless due 
to the weak gauge symmetry or the conservation of the
lepton number $(L)$ and therefore the observed neutrino masses imply
the existence of the new scale -- the only well-known 
scales,~\ie\ weak, strong and
Planck scales, alone are not able to provided this scale.
A very suggestive mechanism is the the see-saw 
mechanism \cite{see-saw}. 

In our extended Anti-GUT \cite{NT} model the see-saw mechanism 
is already built in as far as we assume the existence of the 
three right-handed neutrinos -- Majorana neutrinos -- in the 
range of a new scale in the order of $10^{12}~\GeV$, and the 
predictions of this model are very successful 
for the small mixing angle MSW \cite{MSW} (SMA-MSW) scenario.

Sakharov \cite{sakharov} has pointed out 
that a matter-anti-matter asymmetry can be dynamically 
generated in an expanding Universe if the particle 
interactions and the cosmological evolution satisfy 
the three conditions: {\it (1)} baryon number violation, 
{\it (2)} $C$ and $\CP$ violation, {\it (3)} departure 
from thermal equilibrium. The Anti-GUT model is a natural 
extension model of the 
SM in which $C$ and $\CP$ are 
already violated,~\ie\ this model satisfies if combined with the 
standard cosmology all these Sakharov's conditions, therefore 
the baryon number asymmetry 
should be predictable with this model in the scenario that 
the three Majorana neutrinos 
are very heavy and that they violate the lepton number conservation 
during their out-of-equilibrium decays in the early stage of the 
Universe: Baryogenesis via lepton number violation \cite{FY}.

The presently ``observed'' baryon asymmetry, the baryon 
density-to-entropy density ratio of the Universe,
\begin{equation}
  \label{eq:bau}
  Y_B \equiv\frac{n_B - n_{\bar B}}{s} = (1.0 - 10) \times 10^{-11}\nn,
\end{equation}%
is with the Universe evolution  explained as consequence of the 
spectrum and interactions of the particles which break 
the linear combination of the baryon number ($B$) and lepton number, 
$B-L$, global symmetry at high temperature.

In the following section, we should review briefly the 
Anti-GUT model to calculate the baryogenesis. Then, in 
the next section, we shall discuss the scenario of 
the baryogenesis via lepton number violation. In section 
$4$ we present the results of our model. Section $5$ contains our 
conclusion and resum{\'e}.

%%%%%%%%%%%%%%%%%%%%%%%%%%%%%%%%%%%%%%%%%%%%%%%%%%%%%%%
\section{The extended Anti-GUT model}
\begin{table}[!b]
\caption{All $U(1)$ quantum charges in extended Anti-GUT model.}
\vspace{3mm}
\label{Table1}
\begin{center}
\begin{tabular}{ccccccc} \hline\hline
& $SMG_1$& $SMG_2$ & $SMG_3$ & $U_{\sss B-L,1}$ & $U_{\sss B-L,2}$ & $U_{\sss B-L,3}$ \\ \hline
$u_L,d_L$ &  $\frac{1}{6}$ & $0$ & $0$ & $\frac{1}{3}$ & $0$ & $0$ \\
$u_R$ &  $\frac{2}{3}$ & $0$ & $0$ & $\frac{1}{3}$ & $0$ & $0$ \\
$d_R$ & $-\frac{1}{3}$ & $0$ & $0$ & $\frac{1}{3}$ & $0$ & $0$ \\
$e_L, \nu_{e_{\sss L}}$ & $-\frac{1}{2}$ & $0$ & $0$ & $-1$ & $0$ & $0$ \\
$e_R$ & $-1$ & $0$ & $0$ & $-1$ & $0$ & $0$ \\
$\nu_{e_{\sss R}}$ &  $0$ & $0$ & $0$ & $-1$ & $0$ & $0$ \\ \hline
$c_L,s_L$ & $0$ & $\frac{1}{6}$ & $0$ & $0$ & $\frac{1}{3}$ & $0$ \\
$c_R$ &  $0$ & $\frac{2}{3}$ & $0$ & $0$ & $\frac{1}{3}$ & $0$ \\
$s_R$ & $0$ & $-\frac{1}{3}$ & $0$ & $0$ & $\frac{1}{3}$ & $0$\\
$\mu_L, \nu_{\mu_{\sss L}}$ & $0$ & $-\frac{1}{2}$ & $0$ & $0$ & $-1$ & $0$\\
$\mu_R$ & $0$ & $-1$ & $0$ & $0$  & $-1$ & $0$ \\
$\nu_{\mu_{\sss R}}$ &  $0$ & $0$ & $0$ & $0$ & $-1$ & $0$ \\ \hline
$t_L,b_L$ & $0$ & $0$ & $\frac{1}{6}$ & $0$ & $0$ & $\frac{1}{3}$ \\
$t_R$ &  $0$ & $0$ & $\frac{2}{3}$ & $0$ & $0$ & $\frac{1}{3}$ \\
$b_R$ & $0$ & $0$ & $-\frac{1}{3}$ & $0$ & $0$ & $\frac{1}{3}$\\
$\tau_L, \nu_{\tau_{\sss L}}$ & $0$ & $0$ & $-\frac{1}{2}$ & $0$ & $0$ & $-1$\\
$\tau_R$ & $0$ & $0$ & $-1$ & $0$ & $0$ & $-1$\\
$\nu_{\tau_{\sss R}}$ &  $0$ & $0$ & $0$ & $0$ & $0$ & $-1$ \\ \hline \hline
$\phi_{\sss WS}$ & $\frac{1}{6}$ & $\frac{1}{2}$ & $-\frac{1}{6}$ & $-\frac{2}{3}$ & $1$ & $-\frac{1}{3}$ \\
$S$ & $\frac{1}{6}$ & $-\frac{1}{6}$ & $0$ & $-\frac{2}{3}$ & $\frac{2}{3}$ & $0$ \\
$W$ & $-\frac{1}{6}$ & $-\frac{1}{3}$ & $\frac{1}{2}$ & $\frac{2}{3}$ & $-1$ & $\frac{1}{3}$ \\
$\xi$ & $\frac{1}{3}$ & $-\frac{1}{3}$ & $0$ & $-\frac{1}{3}$ & $\frac{1}{3}$ & $0$ \\
$T$ & $0$ & $-\frac{1}{6}$ & $\frac{1}{6}$ & $0$ & $0$ & $0$ \\
$\chi$ & $0$ & $0$ & $0$ & $0$ & $-1$ & $1$ \\
$\phi_{\sss B-L}$ & $-\frac{1}{6}$ & $\frac{1}{6}$ & $0$ & $\frac{2}{3}$ & 
$-\frac{2}{3}$ & $2$\\ \hline
\hline
\end{tabular}
\end{center}
\end{table}
\noindent

In this section we shall review briefly the extended 
Anti-GUT model \cite{NT}. The extended Anti-GUT model
is based on a large gauge group which is 
the Cartesian product of family specific gauge groups, namely,
\begin{equation}
  \label{eq:AGUTgauge}
  \AGUT \nn,
\end{equation}
where $SMG_i$ denotes $SU(3)_i\times SU(2)_i\times U(1)_i$ (SM gauge group),
and $i$ denotes the generation, \ie\ each ``proto-family'' has 
a certain subgroup of the grand unification group, $SO(10)$. This 
group in Eq.~(\ref{eq:AGUTgauge}) consist \underline{only} 
of those representations that do not mix 
the different irreducible representation of the SM and is spontaneously
breaking down to the $SU(3)\times SU(2)\times U(1)\times U(1)_{B-L}$
at the scale about $1$ to $2$ orders of magnitude under the Planck scale.
The breaking is supposed to occur by six Higgs fields which we have 
invented and denoted by the symbols
$S$, $W$, $T$, $\xi$, $\chi$ and $\phi_{B-L}$ and their quantum 
numbers are given in Table $1$. Finally the breaking of $SU(2)\times U(1)$
of the SM is broken by Weinberg-Salam Higgs field, $\phi_{\sss WS}$. 
(also its quantum numbers are found in Table $1$.)

We summarise here the vacuum expectation values (VEV) of the 
seven Higgs fields which the model contains: %
\begin{list}{\it\arabic{line})}{\usecounter{line}}
\item The smallest VEV Higgs field is the Standard Model 
Weinberg-Salam Higgs field, $\phi_{\sss\rm WS}$,
with the VEV at the weak scale being $246~\GeV/\sqrt{2}$.
\item The next smallest VEV Higgs field is also alone 
in its class and breaks the common $B-L$ gauge group 
$U(1)_{B-L}$, common to the all the families. This symmetry is supposed 
to be broken (Higgsed) at the see-saw scale as needed for 
fitting the over all neutrino oscillation 
scale. This VEV is of the order of $10^{12}~\GeV$ and called 
$\phi_{\sss B-L}$. 
\item The next $4$ Higgs fields are called $\xi$, $T$, $W$, 
and $\chi$ and have VEVs of the order of a 
factor $10$ to $50$ under the Planck unit. That means that if intermediate 
propagators have scales given by the Planck scale, as we assume, they 
will give rise to suppression factors of the order $1/10$ each
time they are needed to cause a transition. 
\item The last one, with VEV of the same order of the 
Planck scale, is the Higgs field $S$, which gives little 
suppression when it is applied, of the order of a factor $1/\sqrt{2}$.
\end{list}

The quantum numbers of the $45$ well-known Weyl particles and 
additional three particles - Majorana neutrinos - are gotten
from the requirement that all anomalies evolving 
$U(1)_{B-L,1}$, $U(1)_{B-L,2}$, $U(1)_{B-L,3}$ vanish strongly 
even without using Green-Schwarz anomaly cancellation mechanism \cite{GS},
\ie\ the extended Anti-GUT model is an anomaly free model. These
quantum numbers are also shown in Table $1$.

Now we can write down the mass matrices which are necessary to
discuss the mechanism of baryogenesis in the following sections: 
the Dirac neutrino mass matrix and the Majorana neutrino 
mass matrix. These matrix elements were gotten using the
technical corrections, factorial factor corrections, 
$\sqrt{\#\,{\rm diagrams}}$ multiplying the mass matrix elements
which take into account the possibilities of permuting the 
contributing Higgs fields. With this technical correction and 
the quantum charges of the Higgs fields the mass matrices,
Dirac neutrino and the Majorana neutrino, are given by: 
\begin{eqnarray}
M^D_\nu \!&\sim&\! \frac{\sVEV{\phi_{\sss WS}}}{\sqrt{2}} \left (\hspace{-0.2 cm}\begin{array}{ccc}
6 \sqrt{35}\,S\,W\,T^2\,\xi^2 & 60 \sqrt{14}\,S^3\,W\,T^2\,\xi^3 
& 60\sqrt{154}\,S^3\,W\,T^2\,\xi^3\,\chi\\
6 \sqrt{35}\,S^2\,W\,T^2\,\xi & 2\sqrt{3}\,W\,T^2 & 2\sqrt{15}\,W\,T^2\,\chi \\
6\sqrt{70}\,S^2\,W\,T\,\xi\,\chi & 2\sqrt{6} W\,T\chi & \sqrt{6} W\,T
\end{array} \hspace{-0.2 cm}\right )\label{eq:diracmass}\\
M_R \!&\sim&\!\sVEV{\phi_{\sss\rm B-L}}\hspace{-0.1cm}
\left (\hspace{-0.2 cm}\begin{array}{ccc}
2\sqrt{210} S^{3}\chi^2\xi^2 & \sqrt{15}S\chi^2\xi & \sqrt{6}S\chi \xi  \\
\sqrt{15} S\chi^2\xi & \sqrt{6} S\chi^2 & \sqrt{\frac{3}{2}} S\chi \\
\sqrt{6} S\chi\xi & \sqrt{\frac{3}{2}} S\chi & S
\end{array} \hspace{-0.2 cm}\right ) \label{RHN}\nn.
\end{eqnarray}

We know neither the Yukawa couplings nor the precise 
masses of the fundamental fermions, but it is a basic 
assumption of the naturalness of our model that these 
couplings are of order unity and random complex
in the Planck scale. In the numerical 
evaluation of the consequences of the model we explicitly take 
into account these uncertain factors of order unity by 
providing each matrix element with an explicit random number 
$\lambda_{ij}$ - with a distribution so that its average 
$\sVEV{\log {\lambda_{ij}}}\approx 0$ and its spreading is 
$64 \%$. Note that the random complex order of unity factors
which are supposed to multiply all the mass matrix elements are
not presented here.
                                              
\section{Fukugita and Yanagida scenario for the lepton number production}
\indent

The weak $SU(2)$ instantons \cite{'tHooft} - 
sphaleron \cite{sphaleron} - guaranteed the rapid
exchanges of the baryon number and lepton number in which 
though $B-L$ is conserved in the time of big bang, when the 
temperature was above the weak scale. But in our model we have the 
three right-handed neutrinos 
decaying in the $L$ quantum number violating way, 
in fact also $B-L$ violating way, at the 
time scale of the see-saw neutrinos. Therefore 
the baryon number violation condition 
of the first Sakharov's condition was effectively fulfilled. 

The assumption in our model that all the coupling constants
and coefficients are of {\em order of unity} and {\em random} 
at the Planck scale, especially having random phases as far as
allowed, implies not only $C$ violation but also $\CP$ violation.

Finally the third condition among the Sakharov conditions --
out-of-equilibrium -- comes about during the Hubble expansion 
due to the excess of the three type of the right-handed Majorana 
neutrinos caused by their masses. From these statements our model is 
seen to implement the scenario of Fukugita and Yanagida \cite{FY}.

In the scenario favoured by our model the \underline{heaviest} 
one among the three 
right-handed neutrinos turns out to gives the dominant contribution 
to the baryon or rather $(B-L)$ quantum produced with the second 
becoming almost same order magnitude. Also it turns out 
that the average lifetime of this heaviest right-handed neutrino is 
of the same order as the Hubble expansion time so that we can count 
that a major part of these right-handed neutrinos first decay 
after inverse decays have essentially stopped. We shall justify 
and discuss these features of the scenario induced by our model 
in the next subsection. But first we shall describe the appearance 
of the baryon asymmetry taking for granted the mentioned 
assumptions so that we {\it (1)} use only the heaviest right-handed 
neutrino and {\it (2)} assume it to live relatively long compared 
to the time needed before $B-L$ is effectively conserved again. 
Really as we shall discuss more below, it is not the full $B-L$ 
which is the important quantity but a special roughly speaking 
``third generation $B-L$'' that is the sufficiently well conserved 
charge quantity. 

\subsection{Conservation of the $B-L$-quantum charge}
 
{\it A priori} the excess of $(B-L)$ quantum number risk 
to be diluted or washed out before the ``accidental'' 
$(B-L)$ conservation of the SM sets in. It is therefore 
very important to argue for that such wash out does not take place.  

An important point is that it actually will turn out that 
if we only thought in terms of $(B-L)$ one would estimate 
an appreciable wash out. However, {\it we define a new ``new charge'' 
$\Hat{(B-L)_3}$ which}\footnote{Note that this $\Hat{(B-L)_3}$ is 
not exactly the same as the $(B-L)_3$ in our model, the latter a 
proto $(B-L)_3$ while $\Hat{(B-L)_3}$ deviates by mixing angles 
from the first one.} {\it is washed away much more slowly.}
In the era until the lightest right-handed neutrino 
has become so hard to produce that there are basically 
no more inverse decay processes producing it going on 
and also $2$-by-$2$ scatterings are supposed negligible, 
we have effective conservation of the following charge 
$\Hat{(B-L)_3}$: we define $\Hat{(B-L)_3}$ to the baryon 
number minus the lepton 
number sitting on those leptons (or quarks) which are 
capable by collision with a Weinberg-Salam Higgs particle 
to produce, in resonance say, the heaviest of the 
right-handed neutrino, but not the two lighter ones. 
 
%In the era after the heaviest right-handed neutrino has gone 
%out-of-equilibrium and essentially has disappeared from the 
%soup there is to a good approximation conservation 
%of the just defined charge $\Hat{(B-L)_3}$ because the main 
%$(B-L)$ violating processes are the production of right-handed 
%Majorana neutrinos followed by their decays which now can only be 
%the two lightest ones. 
%So those leptons (or quarks) that can not produce the two lighter 
%Majorana neutrinos do not participate in $(B-L)$ violation 
%processes and thus we have $(B-L)$-quantum number sitting on them 
%conserved.  

It the era when we can ignore diagrams involving the heaviest see-saw
neutrino a ``third generation'' quark or lepton, \ie\ not coupling to 
the vertices  $N_2\,\phi_{\sss WS}\, \ell$ or 
$N_1\,\phi_{\sss WS}\, \ell$ cannot get converted 
in a $(B-L)$-violating way because the diagrams 
have to contain such vertices.

The protection against dilution by $N_2$ and $N_1$ effects of the 
by $N_3$ produced $(B-L)$ hoped for is thus not relying on our 
model having a gauged $(B-L)_3$ but is a more general mechanism.

The question of whether the $(B-L)$ quantum number produced 
in excess by the decays of the heaviest Majorana neutrino 
will be preserved for the future is thus the question of 
whether the temperature falls so deep that this right-handed 
heaviest neutrino itself gets so hard to produce that we can 
ignore its inverse decay before all these heaviest right-handed 
neutrinos have decayed except for a fraction of order unity.  
The time needed to make the heaviest right-handed neutrino 
effectively unproduceable is the Hubble time corresponding 
to the temperature being equal to the mass of this see-saw neutrino. 
The crucial parameter to settle if this approximation of 
sufficiently slow decay is thus the ratio 
\begin{equation}
  \label{eq:K_3}
K_i\equiv\frac{\Gamma_i}{2 H} \,\Big|_{ T=M_{i} } = \frac{M_{\rm
Planck}}{1.66 \sVEV{\phi_{\sss WS}}^2  8 \pi 
g_{*\,i}^{1/2}}\frac{((M_\nu^D)^{\dagger} M_\nu^D)_{ii}}{M_{i}} \qquad
(i=1, 2, 3)\nn, \end{equation}%
where $\Gamma_i$ is the width of the flavour $i$ Majorana neutrino,
$M_i$ is its mass and $g_{*\,i}$ is the number of the degree of freedom
at temperature $M_i$ (see Eq.~$(7)$). 

%%%%%%%%%%%%%%%%%%%%%%%%%%%%%%%%%%%%%%%%%%%%%%%%%%%%%%%
\subsection{Baryogenesis and $CP$ violation}
\indent

Now a right-handed neutrino, $N_R$, decay into a Weinberg-Salam Higgs 
particle and a left-handed lepton or into the $\CP$ 
conjugate channel.  These two channels have different 
lepton numbers $\pm1$. But, because of our random 
\underline{complex} couplings, the partial widths 
do not have to be the same to next-to-leading-order perturbation theory. 
Defining the measure $\epsilon_i$ 
for the $\CP$ violation in the decay of the right-handed neutrino 
\begin{equation}
  \label{eq:epsilonCP}
 \epsilon_i \equiv{\Gamma_{N_{R_i}\ell}-\Gamma_{N_{R_i}\bar\ell}
\over \Gamma_{N_{R_i}\ell}+ \Gamma_{N_{R_i}\bar\ell}}\nn, 
\end{equation}
where $\Gamma_{N_{R_i}\ell}\equiv \sum_{\alpha,\beta}\Gamma(N_{R_i}
\to \ell^\alpha\phi_{\sss WS}^\beta)$ and 
$\Gamma_{N_{R_i}\bar\ell}\equiv \sum_{\alpha,\beta}\Gamma(N_{R_i}\to
\bar\ell^\alpha \phi_{\sss WS}^{\beta \dagger})$ are the $N_{R_i}$
decay rates (in the $N_{R_i}$ rest frame), summed over the
neutral and charged leptons (and Weinberg-Salam Higgs fields) 
which appear as final states in the $N_{R_i}$ decays one sees that the 
excess of leptons over anti-leptons produced in the decay
of one $N_{R_i}$ is just $\epsilon_i$.

At high temperature ($T\sgeq M_3$) equilibrium
there were as many Majorana neutrinos per species as massless 
fermions, SM fermions. In the case ($T\sgeq M_3$) 
all these Majorana neutrinos decay after the $B-L$ violation 
is switched off.

To be able to calculate Baryogenesis we need to obtain
the total number of effectively massless degree of
freedom of the plasma, $g_{*\,i}$, at temperature of the 
order of lightest right-handed neutrino (about $10^6~\GeV$), $\ie$,
there are $14$ bosons and $45$ well-known Weyl fermions plus
$i$ Majorana particles:
\begin{eqnarray}
  \label{eq:gstarAGUT}
g_{*\,i} \!\! &=&\!\! \sum_{j={\rm bosons}}g_{\,j}\left(\frac{T_j}{T}\right)^4+
\frac{7}{8}\sum_{j={\rm fermions}}g_{\,j}\left(\frac{T_j}{T}\right)^4 %
\nonumber\\
\!\! &=&\!\! 
\underbrace{28\,+ \,\frac{7}{8}\cdot90}_{\textrm{Standrad Model}}\,
+ \underbrace{\frac{7}{4}\cdot i}_{\textrm{see-saw particles}}\,
\nonumber\\
&=& \left\{ \begin{array}{r@{\quad:\quad}l}
            108.5 & i=1\\110.25 & i=2 \\ 112 & i=3
            \end{array}\right.\nn,
\end{eqnarray}
here $T_j$ denotes the effective temperature of any species $j$.
When we have coupling as at the stage discussed between all the 
particles $T_j=T$. The entropy of Planck radiation with the 
degree of freedom $g_{*\,i}$ is 
\begin{equation}
  \label{eq:entropy}
  s_i = \frac{2 \pi^2 \,g_{*\,i}}{45} \,T^3 \nn.
\end{equation}%

Moreover, we should note here that due to the electroweak sphaleron effect,
the baryon number asymmetry $Y_B$ is related to the lepton number 
asymmetry $Y_L$ by \cite{HT}: %
\begin{eqnarray}
  Y_B= a \,Y_{B-L}= \frac{a}{a-1} \,Y_L \nonumber\\
{\rm with}\,\,\, a=\frac{8N_f+4N_H}{22N_f+13 N_H}\nn, \nonumber
\end{eqnarray}
where $N_f$ is the number of generations and $N_H$ the number of Higgs
doublets, this reads in the SM $a=28/79$.

%If there were no dilution -- wash out of the excess of $B-L$
%-- as in case $K\ll1$ the resulting baryon number asymmetry relative 
%to entropy would be
%\begin{equation}
%  \label{eq:Y_Bi}
%  Y_{B,\,i} = a\, Y_{B-L,\,i} = -\frac{0.15\,\epsilon_i}{g_{*\,i}}\nn.
%\end{equation}

Because $K_i$ is not small, we have to expect a dilution effect 
for which we define the suppression factor $\kappa_i$, \ie\ we define 
$\kappa_i$ so that the resulting relative to entropy density, $s_i$,
baryon number density
\begin{equation}
  \label{eq:finalform}
  Y_B \equiv \left|\,\sum_{i=1}^{3}\kappa_i\,\frac{\epsilon_i}{g_{*\,i}} %
\,\right|\nn.
\end{equation}
A good approximation for $\kappa_i$, the dilution factor,
is inferred from Ref. \cite{KT,API}:
\begin{eqnarray}
%K_i\,\sgeq 10^6:\qquad  \kappa_i&=& - (0.1\,K_i)^{\frac{1}{2}} \exp[-\frac{4}{3}\,(0.1\,K_i)^{\frac{1}{4}}]\\
10 \sleq\, K_i \,\sleq 10^6:\qquad\kappa_i&=&-\frac{0.3}{K_i(\ln K_i)^{\frac{3}{5}}}\\
1 \sleq\, K_i \,\sleq 10:\qquad\kappa_i &=& -\frac{1}{2 \,K_i}\nn,\label{gl:K_i}\\
0 \sleq\, K_i \,\sleq 1:\qquad\kappa_i &=& -\frac{1}{6}\nn.
\end{eqnarray}

Note that these dilution factors -- we are taking it -- contain the 
effect of the sphaleron processes we should use, instead of the 
Eq. $(17)$ and Eq. $(18)$, we took the following interpolating redefined 
dilution factor in the range $0 \sleq\, K_i \,\sleq 10$:
\begin{equation}
  \label{eq:neudi}
  0 \sleq\, K_i \,\sleq 10:\qquad\kappa_i = -\frac{1}{2\, \sqrt{{K_i}^2 + 9}}
\end{equation}

Since this dilution factor is smoother than the in 
the Ref.~\cite{KT} defined one, and due to using order-of-one 
complex random 
factors in calculation of $K_i$'s, especially in our model,
baryogenesis comes mainly from the third Majorana neutrino
decay, \ie\ the calculation of the $K_3$ have to be carefully performed
because of $K_3\approx1$, therefore we should better use the
newly defined smoothed out $\kappa_i$ in Eq. $(19)$.

\subsection{$\CP$ violation in decays of the Majorana neutrinos}
\indent

The total decay rate at the tree level (Fig.~$1\!-\!(a)$) is given by
\begin{equation}
  \label{eq:LOCP}
  \Gamma_{N_i}=\Gamma_{N_i\ell}+\Gamma_{N_i\bar\ell}={((M_\nu^D)^\dagger M_\nu^D)_{ii}\over 4\pi \sVEV{\phi_{\sss WS}}^2}\,M_i \nn.
\end{equation}
%%%%%%%%%%%%%%%%%%%%%%%%%%%%%%%%%%%%%%%%%%%%%%%%%%%%%%%
\begin{figure}[t!]
\vspace{5mm}
  \begin{center}
\unitlength=1mm
\begin{fmffile}{fmffig}
\newenvironment{NHf}
    {\begin{fmfgraph*}(35,30)
        \fmfleft{N}\fmfright{fb,f}
        \fmflabel{$\phi_{\sss WS}$}{fb}
        \fmflabel{$N_i$}{N}\fmflabel{$\ell$}{f}}
   {\end{fmfgraph*}}
  \begin{NHf}
    \fmf{plain}{N,v}\fmf{fermion}{v,f}
    \fmf{dashes}{v,fb}
    \fmfdot{v}
\fmflabel{\raisebox{-50mm}{$\!\!\!\!\!\!\!\!\!\!\!\!(a)$}}{v}
 \end{NHf}
\qquad \qquad
 \begin{NHf}
   \fmf{plain}{N,v1}\fmf{plain,tension=.65,label=$N_j$,l.side=left}{v2,v3}
%   \fmf{fermion,left,tension=.25}{v1,v2}
   \fmf{dashes,right,tension=.25}{v1,v2}
   \fmf{fermion,right,tension=.25}{v2,v1}
   \fmf{fermion,tension=.35}{v3,f}
   \fmf{dashes,tension=.35}{v3,fb}
   \fmfdotn{v}{3}
\fmflabel{\raisebox{-50mm}{$\!\!\!\!\!\!\!\!\!\!\!\!(b)$}}{v3}
 \end{NHf}
\qquad \qquad
 \begin{NHf}
   \fmf{plain,tension=1.4}{N,Nv}%\fmf{boson}{fbv,Nv}
   \fmf{fermion,tension=.75}{fbv,Nv}
   \fmf{dashes,tension=.75}{Nv,fv}
   \fmf{dashes}{fbv,fb}\fmf{fermion}{fv,f}
   \fmffreeze
   \fmf{plain,label=$N_j$,l.side=right}{fbv,fv}\fmfdot{Nv,fbv,fv}
\fmflabel{\raisebox{-14.5mm}{$\!\!\!\!\!\!\!\!\!(c)$}}{fbv}
 \end{NHf}
\end{fmffile}
\vspace{12mm}
\caption{Tree level $(a)$, self-energy $(b)$ and vertex $(c)$ diagrams contributing to heavy Majorana neutrino decays.}
\end{center}
\end{figure}
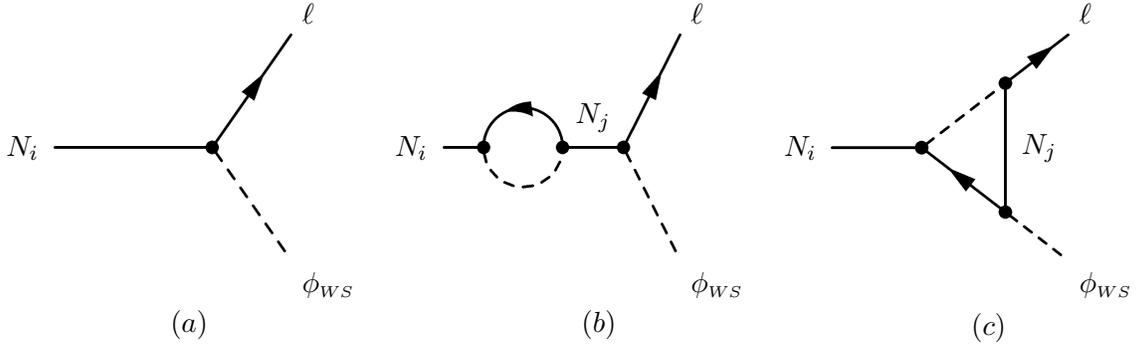
%%%%%%%%%%%%%%%%%%%%%%%%%%%%%%%%%%%%%%%%%%%%%%%%%%%%%%%
The $CP$ violation in the Majorana neutrino decays, 
$\epsilon_i$, arises when the effects of loop are 
taken into account, and at the one-loop, the only $CP$
asymmetry of the vertex contribution comes from the 
diagram shown in Fig.~$1\!-\!(c)$. However there is an other
contribution, from the wave function renormalisation 
(Fig.~$1\!-\!(b)$), which must be also taken into account and which 
gives typically same order amount as the 
vertex one \cite{Luty,BuPlu,CRV}:
\begin{equation}
\label{eq:CPepsilon}
\epsilon_i = \frac{1}{4 \pi \sVEV{\phi_{\sss WS}}^2 ((M_\nu^D)^{\dagger} M_\nu^D)_{ii}}\sum_{j\not= i} {\rm Im}[((M_\nu^D)^{\dagger} M_\nu^D)^2_{ji}] \left[\, f \left( \frac{M_j^2}{M_i^2} \right) + g \left( \frac{M_j^2}{M_i^2} \right)\,\right] 
\end{equation}
where the function, $f(x)$, comes from the one-loop vertex contribution and
the other function, $g(x)$, comes from the self-energy contribution.
These functions can be calculated in perturbation theory 
only for differences between Majorana neutrino masses which 
are sufficiently large compare to its decay widths, \ie\ 
the mass splittings satisfy the condition, 
$\abs{M_i-M_j}\gg\abs{\Gamma_i-\Gamma_j}$:
\begin{eqnarray}
f(x) &=& \sqrt{x} \left[1-(1+x) \ln \frac{1+x}{x}\right]\nn, \label{eq:vas1}\\
g(x) &=& \frac{\sqrt{x}}{1-x} \label{eq:vas2}\nn.
\end{eqnarray}

%%%%%%%%%%%%%%%%%%%%%%%%%%%%%%%%%%%%%%%%%%%%%%%%%%%%%%%

\section{Baryogenesis calculation in Anti-GUT model}

In this section we present the calculation of the baryogenesis 
at first numerically, and then in a crude way by seeking the 
dominating terms. 

The calculation goes by using formula $(\ref{eq:CPepsilon})$, 
$(\ref{eq:vas1})$ and $(\ref{eq:vas2})$ taking as 
the mass matrix $M_\nu^D$, the expression $(3)$, with 
each matrix element further provided with an 
(independent of the other matrix elements) random \underline{complex}
number of order unity. Also the right-handed neutrino 
masses to be used as $M_i~(i=1,2,3)$ in 
the functions $f$ and $g$ in formulas 
$(\ref{eq:vas1})$ and $(\ref{eq:vas2})$ are (in principle)
calculated by inserting random numbers of order unity 
into, in this case, the right-handed neutrino mass matrix 
$(\ref{RHN})$, in the way that each matrix element is again 
provided with an order of unity random (complex) factor.
These random order unity coefficients are in the calculation taken as 
complex pseudo-random numbers.

The quantity $\epsilon_i$ is then calculated $100,000$ 
times and its logarithm, $\ln\!|\epsilon_i|$, is 
averaged over the different sets of random numbers. And the 
parameters which are the suppression factors 
$S, W, T, \xi, \chi$ and $\phi_{B-L}$ are taken for 
the best cases of our previous article fitting
the neutrino masses and its mixing angles:
\begin{eqnarray}
\label{eq:VEVS}
&&\sVEV{\phi_{\sss WS}}=\frac{246}{\sqrt{2}}~\GeV\nn, \nn
\sVEV{\phi_{\sss B-L}}=2.74\times10^{12}~\GeV\nn, \nn\sVEV{S}=0.721\nn,\nonumber\\
&&\nn\sVEV{W}=0.0945\nn,\nn\sVEV{T}=0.0522\nn,\nn\sVEV{\xi}=0.0331\nn,\nn\sVEV{\chi}=0.0345\nn,
\end{eqnarray}
where the vacuum expectation values, except the Weinberg-Salam
Higgs and $\sVEV{\phi_{\sss B-L}}$, are presented in the Planck unit.

That is to say we really use $S$, $W$, $T$, $\xi$, $\chi$ and $\phi_{B-L}$
from one of the best fits including the factorial corrections to 
the charged lepton and quarks, while $\chi$ and $\phi_{B-L}$ are 
fitted to the neutrino oscillation data. Since our model 
unavoidable predicts the small mixing 
angle MSW solution, our calculation is not meaningful 
unless the SMA solution is right. But recently 
Barbieri and Strumia \cite{Barbieri} have studied the neutrino oscillation 
fit using ``non-standard analogies'' method and have shown
that the region of the SMA-MSW solution should be shifted in the 
direction of smaller mixing angle, so that it escapes the day-night
effect exclusion by the Super-Kamiokande measurements. 

\subsection{Results}
\indent

Our total baryogenesis is a sum of three 
contributions, one from each see-saw particle. The total baryon 
density-to-entropy density ratio is given by
\begin{equation}
  \label{eq:sumYB}
  Y_B \equiv\left|\,\sum_{i=1}^{3} Y_{B,\,i} \,\right| = \left|\, \sum_{i=1}^{3} \kappa_i\, \frac{\epsilon_i}{{g_{*}}_i} \,\right|  \nn,
\end{equation}
where $Y_{B,\,i}$ is the contribution due to the decay see-saw neutrino
number $i$ counted upwards in mass.

Dominantly the signs of $\epsilon_3$ and $\epsilon_2$ are strongly 
correlated but since the $\epsilon_3$-contribution is diluted 
away it does not matter and the danger of cancellation is not there.
Actually we included in the numerical calculation the correlation 
of signs effects correctly. In our case the contribution of 
$Y_{B,\,3}$ is dominant and the other contributions are negligible 
compared to $Y_{B,\,3}$. This makes the scheme opposite to 
SUSY constraint ones. 

The numerical results of our best fitting case gives
\begin{eqnarray}
\abs{\epsilon_3} &=& 6.8 \times 10^{-9}\nn,  \qquad K_3 =1.06 \label{eq:epsilon3}\\
\abs{\epsilon_2} &=& 6.0 \times 10^{-9}\nn,  \qquad K_2 =4.29\label{eq:epsilon2}\\
\abs{\epsilon_1} &=& 4.8 \times 10^{-10}\nn, \qquad K_1 =19.8\nn.\label{eq:epsilon1}
\end{eqnarray} %

Using the philosophy of letting order unity random 
numbers being given by a Gaussian distribution presented 
in the article \cite{douglas} we estimate the uncertainty 
in the natural exponent for $Y_B$ to be 
$64~\%\cdot\sqrt{7}$ $\approx 150~\%$. With these error estimations we get
\begin{equation}
   Y_B =  1.46{+5.87\atop-1.17}\times10^{-11}\nn.
\end{equation}

\section{Conclusion and resum{\'e}}
%%%%%%%%%%%%%%%%%%%%%%%%%%%%%%%%%%%%%%%%%%%%%%%%%%%%%%%
\begin{table}[!b]
\caption{Typical fit including averaging over ${\cal O}(1)$ factors. 
All quark masses are running masses at $1~\GeV$ except the top quark 
mass which is the pole mass.}
\begin{displaymath}
\begin{array}{c|c|c}
& {\rm Fitted} & {\rm Experimental} \\ \hline \hline
m_u & 3.1 ~\MeV & 4 ~\MeV \\
m_d & 6.6 ~\MeV & 9 ~\MeV \\
m_e & 0.76 ~\MeV & 0.5 ~\MeV \\
m_c & 1.29 ~\GeV & 1.4 ~\GeV \\
m_s & 390 ~\MeV & 200 ~\MeV \\
m_{\mu} & 85 ~\MeV & 105 ~\MeV \\
M_t & 179 ~\GeV & 180 ~\GeV \\
m_b & 7.8 ~\GeV & 6.3 ~\GeV \\
m_{\tau} & 1.29 ~\GeV & 1.78 ~\GeV \\
V_{us} & 0.21 & 0.22 \\
V_{cb} & 0.023 & 0.041 \\
V_{ub} & 0.0050 & 0.0035 \\
J_{\sss CP} & 1.04\times10^{-5} & 2\!-\!3.5\times10^{-5}\\ \hline \hline
\end{array}
\end{displaymath}
\label{MassenmitFF}
\end{table}
%%%%%%%%%%%%%%%%%%%%%%%%%%%%%%%%%%%%%%%%%%%%%%%%%%%%%%%
We have obtained with the cosmological fit,
\begin{equation}
Y_B ~\big|_{\rm experiments}=  (0.1 - 1.0)\times10^{-10}\nn,
  \label{eq:ergebYB}
\end{equation} %
excellently agreeing prediction of the baryon number-to-entropy ratio 
\begin{equation}
   Y_B =  1.46{+5.87\atop-1.17}\times10^{-11}\nn,
\end{equation} %
from our extended Anti-GUT under use of parameters all 
already fit to either the charged lepton and quark spectra 
or to the neutrino oscillations.

Even the ``discrete fitting'' of the precise choice of 
discrete quantum numbers
of the seven Higgs fields $\phi_{\sss WS}$, $S$, $W$, $T$, 
$\xi$, $\chi$ and $\phi_{B-L}$ of our model were fit 
already to the mass and mixing angle data.

It should be remarked that our model
predicts all fermion masses (neutrino mass square difference ratio) 
and their mixing angles, including Jarlskog triangle, using above 
presented seven Higgs VEVs (The quarks and charged lepton mass 
spectra and their mixing angles are presented in Table~$2$.):
\begin{eqnarray}
  \label{eq:bestresults}
\frac{\Delta m_{\odot}^2}{\Delta m_{\rm atm}^2} \!\!&=&\!\! 5.8{+30\atop-5}\times10^{-3}\\
  \tan^2\theta_{\odot}\!\!&=&\!\! 8.3{+21\atop-6}\,\times10^{-4}\\
  \tan^2\theta_{e3}\!\!&=&\!\! 4.3{+11\atop-3}\,\times10^{-4}\\
  \tan^2\theta_{\rm atm}\!\!&=&\!\! 0.97 {+2.5\atop-0.7}
\end{eqnarray} %
\ie\ our predictions are compatible with 
the \underline{MSW-SMA} solar neutrino solution. Because of smallness 
of the Cabibbo angle induces our solar mixing angle to be so
small that it would stress the model drastically to seek to 
fit one of the series of large solar mixing angle fitting 
regions. 

All of the input parameters are, seven Higgs VEVs, already
determined before the calculation of baryogenesis, so in this 
sense the baryon number-to-entropy ratio is pure 
prediction from our model!

The number of measured quantities which are predictable with our 
model, quark and charged lepton masses and
its mixing angles including the Jarlskog triangle area $J_{CP}$ and also
two mass square differences for the neutrino and the three of 
their mixing angles, is $19$. Our model successfully predicted 
all these quantities using only six parameters\footnote{The VEV of 
Weinberg-Salam we have not counted as a parameter because of the
relation to the Fermi constant.}: the genuine number of predicted
parameters is thus $13$. But we have taken into the predictions 
the quantity $\tan^2\theta_{13}$ for which CHOOZ has only an upper bound.

It should, however, be admitted 
that our prediction is {\it only order of magnitude-wise} 
because we needed to use the assumption that all the 
coupling constants at the fundamental scale are of order unity.

\subsection{What do we learn?}
\indent

In the crudest approximation our agreement means that 
the general success of getting good baryon number prediction
from the Fukugita-Yanagida scheme of having $B$ asymmetry 
from the see-saw scale works well with our specific model. This 
very good agreement 
of our prediction, of course, suggests that our 
specific model may carry some truth. It is first of all 
via the $\CP$-violating parameter, $\epsilon$, that 
different detailed models can make their differences felt 
since even a supersymmetric model doubling of $g_*$ 
is only a factor $2$ hardly distinguishable with our 
only order of unity accuracy. 

We expect the $\epsilon$ which is an expression for the 
overall size of the Yukawa couplings -- it is a loop effect 
relative to the tree diagram -- for given mass splitting -- 
for charged leptons and quarks to be sensitive to the number 
of effective charges in the model used for the mass 
protections. Indeed we expect more general suppression and therefore
smaller $\epsilon$ and thus baryon number if we imagined to 
have control over the $\kappa_i$'s -- the bigger the number 
of charge types used.

When our model then gives a good baryon number result 
we should expect that its number of charges species -- 
effectively -- used is roughly right. Since our model 
has in fact under some restrictions the maximal number 
of gauge charges the success of predicting $\epsilon$
suggests that the right model should have a rather large 
number of charges species. It should, however, in this connection 
be reminded that since we have the field $S$ giving only a 
tiny suppression the tree group we have ``effectively''  
used is not the full Anti-GUT one but rather the subgroup 
of it obtained after breaking by $S$, $SU(3)^2\times SU(2)^2\times U(1)^5$.

It should be also be mentioned that it was in our Anti-GUT 
put in, that the representations of the non-abelian subgroups 
$SU(3)$ and $SU(2)$ were given by a rule \footnote{The rule is that 
the representation of the family specific $SU(2)_i$ and 
$SU(3)_i$ gauge groups are obtained as the same ones as a 
quark or lepton in the SM has when its weak hypercharge 
$y/2$ equals $y_i/2$.} from those of the 
abelian, so that the non-abelian ones play no separate 
role. We did not even list the non-abelian representations 
in Table $1$.

Moreover, it should be stressed that our scenario 
is very different from the SUSY model one: in SUSY model or SUGRA
there is the problem that gravitinos survive and cause an 
unacceptable mass density contribution unless their 
production does not occur due to late inflation \cite{Yanagravi}.
This makes it a problem to obtain the $B-L$ from the see-saw 
neutrino decays and more problematic the heavier the right-handed
neutrino used. This is why Buchm{\"u}ller and Pl{\"u}macher 
\cite{BuPlu} have the lightest of the right-handed neutrino 
provide the baryogenesis.

Since our model is a non-SUSY one we have no gravitino problem and
thus we could equally well make use the heavier Majorana neutrinos 
as of only the light one.

%%%%%%%%%%%%%%%%%%%%%%%%%%%%%%%%%%%%%%%%%%%%%%%%%%%%%%%
\section*{Acknowledgements}
We are grateful to W.~Buchm{\"u}ller, C.D.~Froggatt, 
K.~Kainulainen, M.~Pl{\"u}macher and S.~Lola for useful 
discussions. We would like to thank especially 
T.~Yanagida for discussion of the parameter $K$. We wish to thank 
the Theory Division of CERN for the hospitality extended to us
during our visits where part of this work was done.  H.B.N. 
thanks the EU commission for grants SCI-0430-C (TSTS), 
CHRX-CT-94-0621, INTAS-RFBR-95-0567 and INTAS 93-3316(ext). Y.T. 
thanks the Scandinavia-Japan Sasakawa foundation for grants No.00-22.
%%%%%%%%%%%%%%%%%%%%%%%%%%%%%%%%%%%%%%%%%%%%%%%%%%%%%%%

%
%\end{table}
\end{document}